\documentclass[12pt]{article}
\usepackage{amsmath}
\usepackage{amsthm}
\usepackage{float}
\usepackage{amssymb}
\usepackage{graphicx}
\usepackage{natbib} 
\usepackage{url} 
\usepackage{hyperref} 
 
\newtheorem{theorem}{Theorem}[section]

\newcommand{\blind}{0}

\begin{document}

\def\spacingset#1{\renewcommand{\baselinestretch}%
{#1}\small\normalsize} \spacingset{1}


\if0\blind
{
  \title{\bf Autoregressive Mixture Models for Clustering Time Series}
  \author{Benny Ren \\
    Department of Biostatistics, Epidemiology, and Informatics, \\
    University of Pennsylvania, Philadelphia, USA\\
    and \\
    Ian Barnett \\
    Department of Biostatistics, Epidemiology, and Informatics, \\
    University of Pennsylvania, Philadelphia, USA}
  \maketitle
} \fi

\if1\blind
{
  \bigskip
  \bigskip
  \bigskip
  \begin{center}
    {\LARGE\bf Autoregressive Mixture Models for Serial Correlation Clustering of Time Series Data}
\end{center}
  \medskip
} \fi

\bigskip
\begin{abstract}
Clustering time series into similar groups can improve models by combining information across like time series. While there is a well developed body of literature for clustering of time series, these approaches tend to generate clusters independently of model training which can lead to poor model fit. We propose a novel distributed approach that simultaneously clusters and fits autoregression models for groups of similar individuals. We apply a Wishart mixture model so as to cluster individuals while modeling the corresponding autocovariance matrices at the same time. The fitted Wishart scale matrices map to cluster-level autoregressive coefficients through the Yule-Walker equations, fitting robust parsimonious autoregressive mixture models. This approach is able to discern differences in underlying autocorrelation variation of time series in settings with large heterogeneous datasets. We prove consistency of our cluster membership estimator, asymptotic distributions of coefficients and compare our approach against competing methods through simulation as well as by fitting a COVID-19 forecast model.
\end{abstract}

\noindent%
\vfill

\newpage
\spacingset{2} 
\section{Introduction}
\label{intro}

Modern technologies have accelerated the pace at which data is collected, leading to new frontiers in quantitative research. For example, there is a proliferation of wearable devices and smartphones with sensors that continuously capture large multi-modal time series data at an individual level. Clustering, an important concept in data mining, often elucidates latent characteristics of the study population \citep{fokianos2012biological}. However, these large and often sensitive datasets are well suited for distributed analysis with considerations for privacy  \citep{allard2016new,hong2013survey,yang2019federated,jordan2018communication}. This problem can be addressed by combining low dimensional representations of similar individuals within a sample of heterogenous time series data.

Time series clustering is a well studied topic, see \cite{liao2005clustering} and \cite{maharaj2019time} for a comprehensive review. Time series clustering can be divided into two classes: hard or crisp clustering, where individuals are assigned to a single group and soft or fuzzy clustering, where individuals are assigned to multiple groups with membership weights. Hard clustering commonly follows the procedure of deriving a distance between each pair of time series and performing a hierarchical cluster analysis \citep{montero2014tsclust}. Alternatives to hierarchical cluster analysis include using community network detection and quasi U-statistics \citep{ferreira2016time,valk2012time}. Soft clustering includes centroid based techniques and mixture models and can identify useful mixed memberships which is often preferred over hard clustering in situations with heterogeneous data \citep{d2009autocorrelation,genolini2015kml,wang2015large}.

Time series clustering can concurrently be divided into three types: observation, model and feature based methods. Observation based clustering uses comparison between complete time series such as the many variations of dynamic time warping (DTW), which may be difficult in a large data setting \citep{berndt1994using,paparrizos2015k,cuturi2017soft}. Model based clustering assumes underlying models to derive similarities between time series \citep{coke2010random,piccolo1990distance,wang2019clustering,xiong2002mixtures,gao2020regularized}. Model based clustering is powerful but requires the correct model specification and can be computationally expensive in large datasets. Feature based clustering uses statistics representing each individual time series for clustering, such as the autocorrelation approaches of \cite{galeano2001multivariate}; and \cite{d2009autocorrelation}. Another popular class feature based methods revolve around spectral densities \citep{euan2018spectral,chen2020collective}. Feature based methods work well with large datasets, by efficiently sharing relevant statistics across individuals, thereby bypassing computations that require all data points. However, an important discussion in time series clustering revolves around the role of heterogeneity or noise in the study population, often exhibited through different sample sizes and variances across individuals. Normalization and preprocessing of features are often used in clustering but sensitive in sparse data settings and heterogeneous characteristics should be incorporated into clustering algorithms. Of note, existing methods perform model fitting independently from clustering which is a lost opportunity and can lead to poorer model fit.

We propose a new method that uses a Wishart mixture model (WMM) to address this problem and improve model fit by simultaneously modeling autocovariances along with clustering. The Wishart distribution has many applications in stochastic processes and is closely related to the Gaussian distribution \citep{gourieroux2009wishart,wilson2010generalised}. Under mild conditions, stationary time series also have arbitrarily close causal autoregressive (AR) approximations, (see Corollary 4.4.2 of \cite{brockwell1991time} for more details) lending themselves to techniques that are based in Gaussianity \citep{gupta2013convergence,broersen2000facts}. The Wishart distribution conveniently evaluates scatter matrices (an alternative form of the autocovariance matrix) by their proportionality which alleviates the need to normalize data and incorporates sample size as the degrees of freedom parameter. Individuals in a population may exhibit similar autocovariances in their stationary distribution or weak (second order) stationarity conditions, which can be exploited for clustering. The Wishart mixture model, defined in \cite{hidot2010expectation}, can be used to cluster individuals by their scatter matrices, while simultaneously estimating group scale matrices, making WMM clustering applicable to a wide range of stationary time series.

There's a natural connection between Wishart distributions and AR processes which are well behaved Gaussian processes under certain assumptions. Yule-Walker (YW) coefficients are also conveniently derived from the autocovariance matrix, or any proportional matrices such as the scatter matrix \citep{yule1927vii,walker1931periodicity}. YW estimation is also consistent for causal AR processes \citep{brockwell1991time} and as a result the WMM works well with YW estimation because it clusters matrices by their proportionality while accounting for heterogeneity in the degrees of freedom and variance of the innovations during clustering. We can then use the respective estimated group scale matrices to consistently estimate group specific AR coefficients using the YW equations. This derivation of AR estimates has the advantage of using pooled information across individuals, resulting in a robust and computationally inexpensive estimating procedure, making our method well suited to large datasets. As method of moment estimators, when YW is combined with the mixture approach of the WMM we obtain a mixture of marginal models with familiar asymptotic distributions \citep{liang1986longitudinal,hansen1982large,rosen2000mixtures}. 
At its core, our WMM algorithm is a feature based method to cluster stationary time series by their second-order moments, but under the correct modeling assumption, has the additional benefit estimating an autoregressive mixture model (ARMM). 

In Section \ref{methods}, we derive the WMM model under a variety of parametric assumptions and we detail an EM algorithm to estimate model parameters in each case. Next, we estimate the ARMM using the results from the WMM and outline a model selection procedure as well as detail competing methods. In Section \ref{results}, we compare the WMM with competing approaches through simulation studies and an application with COVID-19 case rate data. Ultimately we find WMM to be a powerful approach for clustering of time series data.

\section{Methods}
\label{methods}
\subsection{Autoregressive Process and Notation}

Our data consist of $I$ independent individuals, with time series from a zero-mean causal AR process,
\small
\begin{align*}
\varepsilon^{(t)}_i =
Y^{(t)}_{i} - \phi^{(1)} Y^{(t-1)}_{i} - \cdots - \phi^{(p)} Y^{(t-p)}_{i} = \left( 1 - \phi^{(1)} B - \cdots - \phi^{(p)} B^p \right) Y^{(t)}_i = \prod_{j=1}^{p}\left(1-\alpha^{(j)} B\right) Y^{(t)}_i
\end{align*}
\normalsize
where innovations are independent and normally distributed as $\varepsilon^{(t)}_i \sim \mathrm{N} (0,\sigma^2_i)$ and the lag operator is denoted as $B$. Innovation variances $\sigma_i^{2}$ are specific to individual $i$ and assumed to be finite. Individual $i$ has a time series vector of length $n_i$ and we denote the time point index through the superscript in parentheses, $\mathbf{y}_i = \left\{ y_i^{(1)}, y_i^{(2)}, y_i^{(3)},\dots, y_i^{(t)},\dots, y_i^{(n_i)} \right\}$. Along with causality, we also make a stationarity assumption. After some algebra with partial fractions, the process can be written as a causal sequence of $\varepsilon^{(t)}_i$
\begin{align*}
Y^{(t)}_i = \sum_{j=1}^{p}\left[\frac{c_j}{1-\alpha^{(j)} B}\right] \varepsilon^{(t)}_i = \sum_{j=0}^{\infty} a^{(j)} \varepsilon^{(t-j)}_i
\end{align*}
where $c_j$ are constants from partial-fraction decomposition in order to obtain a sum of geometric series. The causal sequence can describe many stochastic processes such as the autoregressive moving average (ARMA) family. Under the stationarity assumption we have autocovariances which only depends on lag $k$:
$$
\gamma_i(k) = \sigma^2_i \sum_{j=0}^{\infty} a^{(j+k)} a^{(j)} = \sigma^2_i \tau(k)
$$
where $\tau(k)=\sum_{j=0}^{\infty} a^{(j+k)} a^{(j)}$ are finite.

A  straightforward result of causality is that $\mathbf{y}_i$ is a Gaussian process where weak and strict stationarity are equivalent. Stationarity results in a marginal multivariate normal distribution for any window of $\mathbf{y}^{(t)}_i = \left[ Y^{(t)}_i, \cdots, Y^{(t+K-1)}_i \right]^T \sim \mathrm{N} (0,\sigma^2_i \zeta)$, with autovariance matrix, $\sigma^2_i \zeta$, that is constant across time, such that $\left[ \zeta \right]_{rc} = \tau(|r-c|)$ where $r$ and $c$ indicate the indices of matrix $\zeta$. As a result, the outer product follows a singular Wishart distribution, $\left[ \mathbf{y}^{(t)}_i \right] \left[ \mathbf{y}^{(t)}_i \right]^T \sim \mathcal{W}_{K}(\sigma^2_i \zeta, 1)$, where AR coefficients are mapped to $\zeta$, leaving innovation variance $\sigma^2_i$ as a scaling term \citep{bodnar2008properties}.

In a study population there may exist a set of coefficients, described by $\zeta$, which can be clustered together to form a population level model for each cluster. The $\sigma^2_i$ allow heterogeneous variances to be incorporated into clustering. We use the Wishart distribution to evaluate $\mathbf{S}_{i} = \sum^{n_i}_{t=1} \left[ \mathbf{y}^{(t)}_i \right] \left[ \mathbf{y}^{(t)}_i \right]^T$, a square $K$ dimension non-singular scatter matrix. Replacing the elements of $\mathbf{S}_i$ using its autocovariance MLE results in a Toeplitz, bi-symmetric structure such that $\left[ \mathbf{S}_i \right]_{rc} = n_i\hat{\gamma}_i(|r-c|)$ where $r$ and $c$ indicate the indices of matrix $\mathbf{S}_i$. We only need to calculate $K$ autocovariance statistics for each individual. In general, our method relies only on second order conditions, matching a time series with matrix $\zeta$, making it applicable to a broad array of stationary processes. 

\subsection{Autoregressive Mixture Model}
The corresponding group AR model our algorithm identifies is given by 
\begin{equation}
y_{i}^{(t)}=\sum_{k=1}^{K-1} {\phi}_{g}^{(k)} y_{i}^{(t-k)}+\varepsilon_{i}^{(t)}, \quad \varepsilon_{i}^{(t)} \sim N\left(0, \sigma_i^{2}\right) \label{AR}
\end{equation}
where $\phi_{g}^{(k)}$ and $\zeta_g$ are the parameters shared by the cluster/group, $g$ denotes the group index, and $k$ is the lag index. Coefficients $\Phi_g=\left[ {\phi}_{g}^{(1)}, {\phi}_{g}^{(2)}, \dots, {\phi}_{g}^{(K-1)}\right]^T$ are mapped to matrix $\zeta_g$ and $\sigma^2_i$ allows for groups to be defined by combinations of different coefficients and innovation variances. Using the YW equations, coefficients can be estimated using any matrix proportional to autocovariance and by extension, anything proportional to $\zeta_g$. Because YW and AR models are well studied topics, there are many tools at our disposal to complement our analysis. For example, assuming a causal and stationary AR process, from Theorem 8.1.1 in \cite{brockwell1991time}, we have: $\frac{1}{n_i} \mathbf{S}_i \stackrel{P}{\rightarrow} \sigma_i^2 \zeta_g$, $\hat{\sigma}_i^{2} \stackrel{P}{\rightarrow} \sigma_i^{2}$, and $\hat{\Phi}_g \stackrel{P}{\rightarrow} \Phi_g$. In addition, \cite{qiu2013efficient} and \cite{shao2011autoregressive} proposed detrending procedures which retains the asymptotic properties of YW. Such a procedure can be applied to each individual time series before fitting our WMM.

\subsubsection{Wishart Mixture Model}
First we define the Wishart mixture model for $G$ total number of groups or clusters,
$$
Z_{i g} \sim \operatorname{Mult}\left(1 ; \pi_{1}, \pi_{2}, \ldots, \pi_{G}\right) \quad 
\mathbf{S}_{i} |\left\{Z_{i g}=1\right\} \sim \mathcal{W}_{K}\left( \Sigma_{g}, n_{i}\right) 
$$
\begin{align*}
f_{W}(\mathbf{S}_i | \Sigma_g, n_i) &= \frac{|\mathbf{S}_i|^{(n_i-K-1) / 2} \exp \left( - \frac{1}{2} \operatorname{trace}\left( \Sigma_g^{-1} \mathbf{S}_i\right)\right) }{2^{n_i K / 2} \pi^{K(K-1) / 4}| \Sigma_g |^{n_i / 2} \prod_{k=1}^{K} \Gamma\left(\frac{n_i-k+1}{2}\right)} \\
&= 
\left[
\frac{\left| \mathbf{S}_{i} \right|^{-(K+1) / 2}}{\pi^{K(K-1) / 4} \prod_{k=1}^{K} \Gamma\left(\frac{n_i-k+1}{2}\right) }
\right]
\left[
\left|\frac{ \mathbf{S}_{i} \Sigma_{g}^{-1}}{2}\right|^{n_i / 2} 
\exp \left( - \operatorname{trace} \left( \frac{ \Sigma_{g}^{-1} \mathbf{S}_{i} }{2} \right) \right)
\right] \\
&= c(\mathbf{S}_i \mid n_i ) h( \mathbf{S}_i \mid \Sigma_g, n_i )
\end{align*}
where  $g \in \{1,2,\dots,G\}$, the missing group indicators $Z_{i g}$ follows a multinomial distribution with $\pi_1, \pi_2, \dots, \pi_G$ as the mixing probability and $\mathbf{S}_{i}$ follows a Wishart distribution with $\Sigma_g= \kappa_g \zeta_g$ as the group scale matrix, $\kappa_g$ is a positive scalar and $n_i$ degrees of freedom. 

The Wishart density is characterized by evaluating $\Sigma_g^{-1} \mathbf{S}_i$ through $h( \mathbf{S}_i \mid \Sigma_g, n_i )$, which relates matrices by their proportionality. For example, if $\frac{1}{n_i} \mathbf{S}_i \stackrel{P}{\rightarrow} \sigma^2_i \zeta_g$, we then have $\Sigma_g^{-1} \mathbf{S}_i \approx n_i \kappa^{-1}_g \sigma^2_i \mathbf{I}$ and the Wishart density is asymptotically driven by $n_i$.

The complete data likelihood for $\Theta=\left\{ \pi_{1}, \ldots, \pi_{g}, \Sigma_{1}, \ldots, \Sigma_{g} \right\}$ is given as 
\begin{equation} \label{LWMM}
L(\Theta)=\prod_{i=1}^{I} \prod_{g=1}^{G}\left(\pi_{g} f_{W}\left(\mathbf{S}_{i} | \Sigma_{g}, n_i\right)\right)^{z_{i g}}
\end{equation}
Using the EM algorithm, we estimate $\pi_{g}$, $\Sigma_{g}$, and impute $z_{i g}$ \citep{dempster1977maximum, hidot2010expectation}. In the estimation step, the function $Q\left( \Theta, \hat{\Theta}^{(t)} \right)$, with current estimate $\hat{\Theta}^{(t)}$, is given as
\begin{align*}
Q\left( \Theta, \hat{\Theta}^{(t)} \right) &= E \left[ \log L(\Theta) | \mathbf{S}_{1}, \mathbf{S}_{2}, \dots, \mathbf{S}_{I}, \hat{\Theta}^{(t)}\right] \\
&= \sum_{i=1}^{I} \sum_{g=1}^{G} 
\Bigg( 
E \left[z_{i g} | \mathbf{S}_{i}, \hat{\Theta}^{(t)}\right] \log \left( {\pi}_{g} f_{W}\left(\mathbf{S}_{i} | {\Sigma}_{g}, {n}_{i} \right)\right)
\Bigg) .
\end{align*}
After conditioning and noting that $\hat{z}_{i g}^{(t+1)} = E \left[z_{i g} | \mathbf{S}_{i}, \hat{\Theta}^{(t)}\right] = \operatorname{Pr} \left( Z_{i g}=1 | \mathbf{S}_{i}, \hat{\Theta}^{(t)} \right)$, we get
\footnotesize
\begin{equation}\label{EM1z}
\hat{z}_{i g}^{(t+1)} 
=
\frac{\operatorname{Pr} \left(Z_{i g}=1\right) f \left(\mathbf{S}_{i} |Z_{i g}=1, \hat{\Sigma}_{g}^{(t)}, {n}_{i} \right)}{f\left(\mathbf{S}_{i} | \hat{\Theta}^{(t)}\right)} 
=
\frac{\hat{\pi}_{g}^{(t)} f_{W}\left(\mathbf{S}_{i} | \hat{\Sigma}_{g}^{(t)}, {n}_{i} \right)}
{\sum_{j=1}^{G} \hat{\pi}_{j}^{(t)} f_{W}\left(\mathbf{S}_{i} | \hat{\Sigma}_{j}^{(t)}, {n}_{i} \right)}
=
\frac{\hat{\pi}_{g}^{(t)} h \left(\mathbf{S}_{i} | \hat{\Sigma}_{g}^{(t)}, {n}_{i} \right)}
{\sum_{j=1}^{G} \hat{\pi}_{j}^{(t)} h \left(\mathbf{S}_{i} | \hat{\Sigma}_{j}^{(t)}, {n}_{i} \right)} 
\end{equation}
\normalsize
and $c(\mathbf{S}_i \mid n_i )$ is constant across all densities. Here $n_i$ and $\sigma^2_i$ controls the extent of influence an individual's noise has on clustering by accessing membership through $h( \mathbf{S}_i \mid \Sigma_g, n_i )$ at different $g$'s. A small $n_i$, often associated with a noisy estimate of $\mathbf{S}_i$, results in a flat density function, leading to mixed soft clustering assignments through equation \eqref{EM1z}. This allows the data from noisy individuals to be dispersed throughout the each group's estimation rather than assumed by any individual group, contrary to hard clustering techniques.

In the maximization step, maximizing $Q\left( \Theta, \hat{\Theta}^{(t)} \right)$ under the constraint \(\sum_{g=1}^{G} \hat{\pi}^{(t+1)}_{g}=1,\) yields our update for the mixing probability \(\hat{\pi}_{g}^{(t+1)}\) as
\begin{equation} \label{EM1pi}
\hat{\pi}_{g}^{(t+1)}=\frac{1}{I} \sum_{i=1}^{I} \hat{z}_{i g}^{(t+1)} .
\end{equation}
The score function for $\Sigma_g$ is
$$
\frac{\partial Q\left(\Theta, \hat{\Theta}^{(t)} \right) }{\partial {\Sigma}_{g}} = 
\sum_{i=1}^{I} \hat{z}_{i g}^{(t+1)} \left(\frac{1}{2} \Sigma_{g}^{-1} \mathbf{S}_{i} \Sigma_{g}^{-1}-\frac{n_{i}}{2} \Sigma_{g}^{-1}\right) =0 
$$
which yields our update as
\begin{equation} \label{EM1sigma}
\hat{\Sigma}_{g}^{(t+1)}=\frac{\sum_{i=1}^{I} \hat{z}_{i g}^{(t+1)} \mathbf{S}_{i}}{ \sum_{i=1}^{I} n_i \hat{z}_{i g}^{(t+1)}} .
\end{equation}
The estimation of $\hat{\Sigma}_g$, which maps to the group AR coefficients, is done simultaneously along with clustering, allowing all individuals to be leveraged for the estimation of each group model under the soft clustering assignments. As $\frac{1}{n_i} \mathbf{S}_i \stackrel{P}{\rightarrow} \sigma^2_i \zeta_g$, then $\hat{\Sigma}_{g} \approx \frac{\sum_{i=1}^{I} n_i {z}_{i g} \sigma^2_i}{ \sum_{i=1}^{I} n_i {z}_{i g} } \zeta_g = \kappa_g \zeta_g$. The $\kappa_g$ are the group mean of variances and if variance is constant within the group, then $\sigma^2_i=\kappa_g$ and $\Sigma_g^{-1} \mathbf{S}_i \approx n_i \mathbf{I}$. This is the ideal scenario for clustering, where the Wishart density is only a function of $n_i$. But if variance is not constant, then $\Sigma_g^{-1} \mathbf{S}_i \approx n_i \kappa^{-1}_g \sigma^2_i \mathbf{I}$, where the ratio of the individual and group mean variance is used to adjust the Wishart density. Together, proportionality, $\sigma^2_i$ and $n_i$ controls the peakedness of the Wishart density and determine each individual's group membership. For example, a high value for $\sigma^2_i$ makes achieving proportionality difficult, requiring a large $n_i$.

Repeating the EM algorithm, equations, \eqref{EM1z}, \eqref{EM1pi}, and \eqref{EM1sigma} until convergence leads to the estimate of $\Theta$. The $\hat{\Sigma}_{g}$ are proportional to $\zeta_g$, making it a valid statistic for YW estimation. The final group indicator integer is imputed as the index that maximizes $\hat{z}_{ig}$, also known as the \textit{maximum a posteriori} (MAP) rule for values $z_{ig}$ such that ${z}_{ig^*}=1$ where $g^* = \underset{g}{\operatorname{argmax}} \left\{ \hat{z}_{ig} \right\}$.

\subsubsection{Variation of the WMM and Effective Degrees of Freedom}

The $\left[ \mathbf{y}^{(t)}_i \right] \left[ \mathbf{y}^{(t)}_i \right]^T$ are marginally distributed as a Wishart distribution. However, $\left[ \mathbf{y}^{(t)}_i \right] \left[ \mathbf{y}^{(t)}_i \right]^T$ and $\left[ \mathbf{y}^{(t+k)}_i \right] \left[ \mathbf{y}^{(t+k)}_i \right]^T$ are not independently distributed. It's has been proposed to account for this correlation by using the effective degrees of freedom. Naturally, we may choose $n_i$ as the effective degrees of freedom as it aligns with moment matching \citep{pivaro2017exact}. Alternatively, we propose to account for correlation within the scatter matrix by scaling the sample size by a positive factor, $\lambda n_i$ \citep{afyouni2019effective,quenouille1947large,bartlett1946theoretical}. In addition, computing Wishart densities can be numerically unstable when $n_i$ is large as the distribution function becomes very peaked, making convergence highly sensitive to initial parameter values.  We address numerical instability and correlation with a modified version of the above proposed EM algorithm where we estimate $\lambda$. 

We propose an extension of the EM algorithm which adjusts degrees of freedom at a cluster level by the scaling with a $\lambda_g$ group adjustment term to $n_i$, where $\Lambda = \{\lambda_1, \lambda_2, \dots, \lambda_G \}$. The update of $\lambda_g$ is calculated by solving score function
$$
\frac{\partial }{\partial \lambda_g } Q\left(\Theta, \Lambda, \hat{\Theta}, \hat{\Lambda} \right) 
=
\sum_{i=1}^{I} \hat{z}_{i g} n_i \log \left|\frac{\mathbf{S}_{i} \hat{\Sigma}_{g}^{-1}}{2}\right|-\sum_{i=1}^{I} \hat{z}_{i g} n_i \sum_{k=1}^{K} \psi\left(\frac{1}{2}\left( {\lambda}_{g}n_{i}-k+1 \right)\right) = 0 .
$$
We update equation \eqref{EM1z} using 
\begin{equation} \label{EM3z}
\hat{z}_{ig}^{(t+1)}
= \frac{ \hat{\pi}^{(t)}_g f_W \left( \mathbf{S}_i | \hat{\Sigma}^{(t)}_g, \hat{\lambda}^{(t)}_g n_i \right) }{
\sum_{j=1}^G \hat{\pi}^{(t)}_j f_W \left( \mathbf{S}_i | \hat{\Sigma}^{(t)}_j, \hat{\lambda}^{(t)}_j n_i \right) }
\end{equation}
and \eqref{EM1sigma} becomes 
\begin{equation} \label{EM3sigma}
\hat{\Sigma}^{(t+1)}_{g} = \frac{ \sum^I_{i=1} \hat{z}^{(t+1)}_{ig} \mathbf{S}_i }{ \hat{\lambda}^{(t)}_g \sum^I_{i=1} n_i \hat{z}^{(t+1)}_{ig}  }.
\end{equation}
We numerically update $\lambda_g$ using
\begin{equation} \label{EM3lambda}
\hat{\lambda}^{(t+1)}_g = \underset{\lambda_g}{\operatorname{argmin}} \left\{ 
\left( 
\frac{\partial }{\partial \lambda_g } Q\left(\Theta, \Lambda, \hat{\Theta}^{(t+1)}, \hat{\Lambda}^{(t)} \right)
\right)^2 
\right\}
\end{equation}
such that $\lambda_g \in \Big( \frac{K-1}{\min(n_i)} ,U \Big]$ and an upper bound prevents numerically instability incurred due to having a large degree of freedom. $\lambda_g$ can be solved using a constrained optimization algorithm such as the constrained Broyden-Fletcher-Goldfarb-Shanno algorithm \citep{byrd1995limited}. For this additional algorithm, repeat, in order, \eqref{EM3z}, \eqref{EM1pi}, \eqref{EM3sigma}, and \eqref{EM3lambda} until convergence. The above versions of the proposed EM algorithm allow for different assumptions on the Wishart distribution, namely with respect to how the degrees of the freedom parameter are handled, and as such we treat both approaches as competing methods.

\subsubsection{Yule-Walker Estimators}

The YW equations can provide an alternative representation of the WMM model as a mixture of marginal models \citep{rosen2000mixtures}. The group estimate, $\hat{\Sigma}_g \approx \kappa_g \zeta_g$ is an proportional estimator of the autocovariance matrix. We can use $\hat{\Sigma}_{g}$ in the YW estimator for AR coefficients as it relies on a matrix that is proportional to the autocovariance matrix. The matrix can be blocked as
$$
\hat{\Sigma}_g = 
\left[\begin{array}{c|c}
q_g & \mathbf{u}_g^T  \\
\hline
\mathbf{u}_g & \mathbf{Q}_g
\end{array}\right] \succ 0
$$
where $q_g$ is scalar, $\mathbf{u}_g$ is a $K-1$ vector, and $\mathbf{Q}_g$ is a $K-1$ square matrix. The YW system for estimating AR coefficients is given as the method of moments estimator $\hat{ \Phi }_g = \left[ \hat{\phi}_{g}^{(1)}, \hat{\phi}_{g}^{(2)}, \dots, \hat{\phi}_{g}^{(K-1)}\right]^T = \mathbf{Q}_g^{-1} \mathbf{u}_g$. The denominator, $\sum^I_{i=1} n_i \hat{z}_{ig}$, of $\hat{\Sigma}_g$ cancel when calculating $\hat{ \Phi }_g$ resulting in a weighted sum of moments from different individuals, akin to a weighted generalized estimating equation (GEE) or generalized method of moments approach \citep{liang1986longitudinal,hansen1982large}. 

The YW estimation results in group AR models given in equation \eqref{AR}. The $\hat{\Sigma}_g$ maps to the autoregressive parameters, $\hat{\Phi}_{g}$ and is estimated using data across individuals, creating a robust group model. Incorporating individual $i$'s group indicator, $z_{ig}$ from the WMM, we arrive at the final ARMM
\begin{equation} \label{ARMM}
y_{i}^{(t)}=\sum_{g=1}^{G} z_{i g} \left[\sum_{k=1}^{K-1} {\phi}_{g}^{(k)} y_{i}^{(t-k)}\right] + {\varepsilon}_{i}^{(t)}, \quad {\varepsilon}_{i}^{(t)} \sim N\left(0, \sigma^2_i \right).
\end{equation}
The $z_{i g}$ govern AR model membership for individual $i$; ${\phi}_{g}^{(k)}$ characterizes the autoregressive behavior; $\sigma^2_i$ allows for heterogeneity within a set of AR coefficients and $\sigma^2_i$ along with $n_i$ is incorporated into the WMM clustering through \eqref{EM1z} and modulates the influence of individual $i$.

\begin{figure}[H]
\centering
\includegraphics[width=6in]{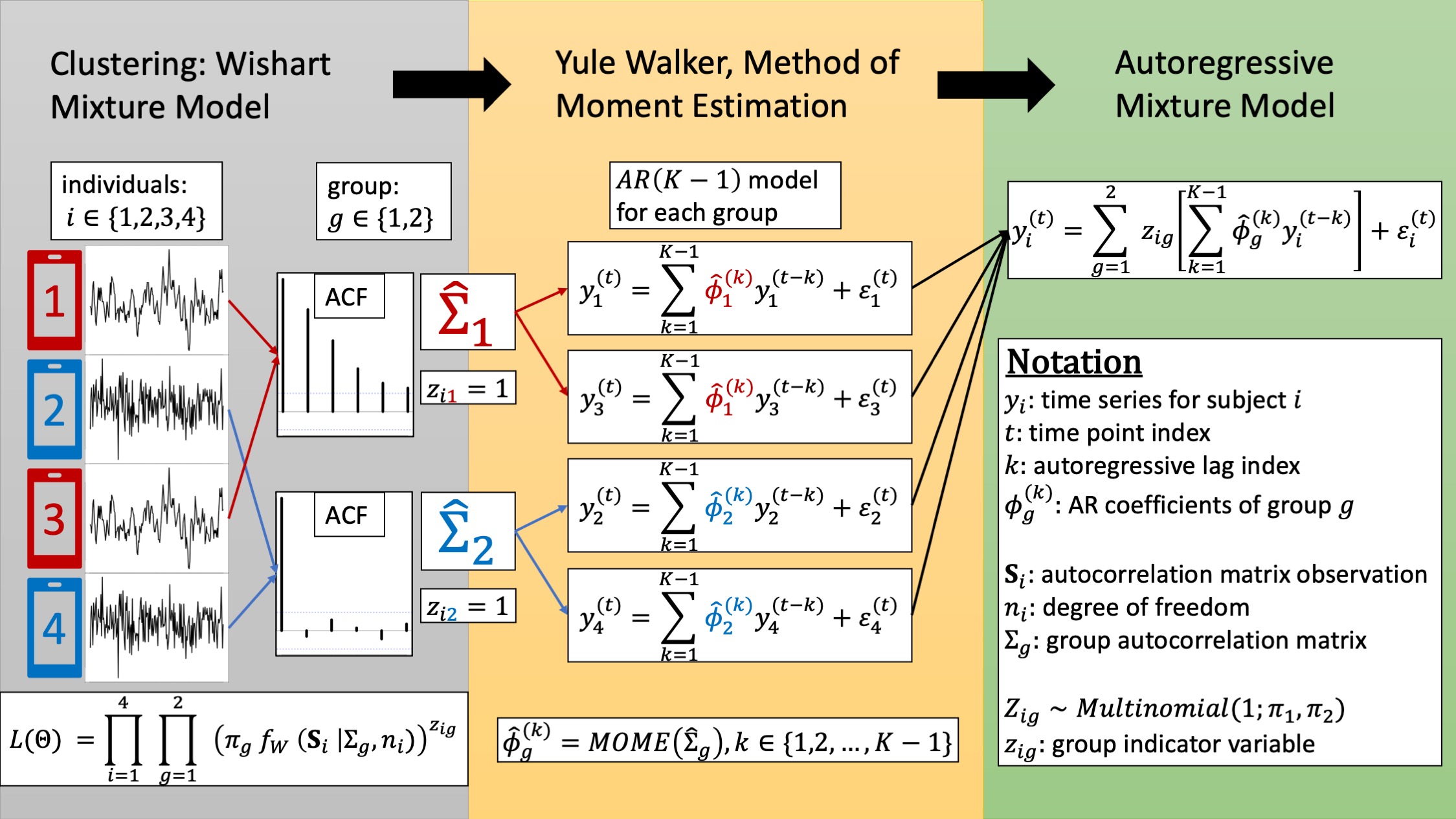}
\caption{ \textbf{Schematic of Parameter Estimation for the Wishart Mixture Model}. In the first phase, four example time series are mapped to scatter matrices, $\mathbf{S}_i$, and are processed by the WMM. In the second phase, $\hat{\Sigma}_g$ is used to estimate the AR coefficients $\Phi_g$ via Yule-Walker estimation. In the third phase, group indicators $z_{ig}$ are combined with the group AR models to create the ARMM.}
\label{fig:1}
\end{figure}

\subsubsection{Selecting the number of clusters ($G$) and the lag ($K-1$)}
Selecting the AR lag \textit{a priori} is an open question; it can be based off of domain knowledge or autocorrelation plots. After estimation, we can also use a penalization criteria such as AIC or BIC to select the lag \citep{akaike1974new,schwarz1978estimating,hurvich1989regression}. However, we recommend selecting a conservative number of lags, greater than the true value of $K$, because the additional coefficient estimates tend to be close to zero and may also be evaluated using its asymptotic distribution. Fitting the WMM improves with number of lags $K$, as our algorithm only depends on the estimation of the stationary autocovariance matrix. First, we may select $G$ based on a selection criterion then address selection of $K$ for each group (see Chapter 6.5 of \cite{madsen2007time} for details on model selection for $K$). At this point, we have estimated MAP group memberships and already calculated all the necessary statistics for estimation and inference of all possible AR($p$) coefficients (where $p \leq K-1$), and $\sigma^2_i$ (and AIC/BIC are functions of $p$, $\hat{\sigma}^2_i$ and $n_i$), so we may use perform selection to reduce the number of lags at specific groups without extra computations involving the complete dataset.

The number of groups $G$ can selected based on AIC or BIC. Using the MAP estimate of $z_{ig}$, the AIC and BIC contribution for an individual is, $n_i \log \hat{\sigma}_{i}^{2}$,
where we use YW estimates, $\hat{\sigma}^2_i = \sum^G_{g=1} z^{(\text{MAP})}_{ig} \hat{\gamma}_i(0) \left[ 1 - \mathbf{u}^T_g \mathbf{Q}_g^{-1} \mathbf{u}_g / q_g \right] > 0$ (inequality by Schur Complement). This estimate also has the benefit of using previously calculated statistics and does not require any new computations. The AIC and BIC for number of groups are
\begin{equation} \label{AIC}
\text{AIC} = 2r + \sum^I_{i=1} n_i \log \hat{\sigma}_{i}^{2}, \quad \text{BIC} = r \log \left( \sum^I_{i=1} n_i \right) + \sum^I_{i=1} n_i \log \hat{\sigma}_{i}^{2}
\end{equation}
where $r = G * K-1 $.

\subsection{Competing Methods}

In our analyses, we consider alternative clustering methods suitable under a large data setting. We explore cluster analysis methods based on retrieving individual level statistics relevant to AR and stationary processes. We use cluster analyses that calculate pairwise distances between all individuals for hierarchical clustering, which are outlined in the \texttt{TSclust R} package \citep{montero2014tsclust}. We compared hierarchical clustering based on several relevant distances. 

First we consider the distances defined by \cite{galeano2001multivariate} between two time series derived using autocorrelations
\begin{align*} 
d_{ACF} \left(\mathbf{y}_{i}, \mathbf{y}_{j}\right)=
\sqrt{\sum_{k=1}^{K-1} \left( \hat{\rho}_i (k) - \hat{\rho}_j (k) \right)^{2}}.
\end{align*}
We also define $d_{PACF} \left(\mathbf{y}_{i}, \mathbf{y}_{j}\right)$, as $d_{ACF} \left(\mathbf{y}_{i}, \mathbf{y}_{j}\right)$, replacing the autocorrelation with partial autocorrelation as our second distance measure. Finally, we use Piccolo distance defined as
$$
d_{PIC}\left(\mathbf{y}_{i}, \mathbf{y}_{j}\right)=
\sqrt{\sum_{k=1}^{K-1}\left( \hat{\phi}^{(k)}_i - \hat{\phi}^{(k)}_j \right)^{2}},
$$
where $\hat{\phi}^{(k)}_i$ are AR coefficients, for our third distance measure \citep{piccolo1990distance}. 

We also propose a soft clustering analog to the Piccolo distance where the AR coefficients are averaged across individuals to obtain the model average. We use a Gaussian mixture model (GMM) and treat individual $i$'s AR coefficients (maximum likelihood estimates), $\mathbf{b}_i = \left[ \hat{\phi}^{(1)}_i, \hat{\phi}^{(2)}_i, \dots, \hat{\phi}^{(K-1)}_i \right]^T$ as observations of a multivariate normal distribution. The GMM complete data likelihood is given as
\begin{align*}
L( \pi_1, \dots, \pi_G, \Phi_1, \dots, \Phi_G, \Omega_1, \dots, \Omega_G ) =
\prod^I_{i=1} \prod^G_{g=1} \left( 
\pi_g 
f \left( 
\mathbf{b}_i | \Phi_g, \Omega_g
\right)
\right)^{z_{ig}}
\end{align*}
where $\mathbf{b}_i | \left\{Z_{i g}=1\right\} \sim N \left( \Phi_g , \Omega_g \right)$. We calculate $\hat{\pi}_g, \hat{\Phi}_g$, and $\hat{z}_{ig}$ using the GMM, rather than the WMM, using \texttt{R} package \texttt{mclust} \citep{fraley2014mclust}. Our WMM methods already serves as a soft clustering analog for autocorrelation distance. 

Finally, we also evaluate the Hierarchical Spectral Merger (HSM) algorithm, \texttt{R} package \texttt{HSMClust}, another method for stationary time series \citep{euan2018hierarchical}. The HSM is based on Total Variation distance of the normalized spectral densities
$$
d_{T V}(f, g)=1-\int \min \{f(\omega), g(\omega)\} \mathrm{d} \omega
$$
where normalization is given as $\widehat{f}(\omega)=\tilde{f}(\omega) / \hat{\gamma}(0)$. Spectral densities are closely related to the causal form of ARMA processes making the HSM and our WMM closely related in their mathematical reasoning. 

\section{Results}
\label{results}

\subsection{Asymptotic Results}

The asymptotic behavior of YW estimation for an individual time series is well studied. Writing the time series in matrix form, $\mathbf{y}_i = \mathbf{X}_i\Phi_i + \boldsymbol{\varepsilon}$ where $\mathbf{X}_i$ is the proper arrangement of $\mathbf{y}_i$, we have the following consistency results.
\begin{theorem} \label{thm1}
(Theorem 8.1.1 from \cite{brockwell1991time})
Assuming $\mathbf{y}_i$ is a causal and stationary AR($K-1$) process, as $n_i \rightarrow \infty$, then $\hat{\sigma}_i^{2} \stackrel{P}{\rightarrow} \sigma_i^{2}$, $\frac{1}{n_i} \mathbf{S}_i \stackrel{P}{\rightarrow} \sigma^2_i \zeta_i$, $\frac{1}{n_i} \mathbf{X}_i^T \mathbf{X}_i \stackrel{P}{\rightarrow} \Omega_i$, $\hat{\Phi}_i \stackrel{P}{\rightarrow} \Phi_i$, and
$$
n_i^{1 / 2}\left(\hat{\Phi}_i - \Phi_i \right) \stackrel{D}{\sim} \mathrm{N}\left(0, \sigma_i^{2} \Omega_i^{-1}\right)
$$
where $\Omega_i$ is the true $K-1$ dimension autocovariance matrix.
\end{theorem}

It is well known that EM converges to the local maximum; typically multiple runs from random
initializations are executed to determine the MLE. We have found that, under certain conditions, the WMM is a consistent estimator as $n_i$ increases i.e. as more observations per individual are obtained:

\begin{theorem} \label{thm2}
Assuming $\mathbf{y}_i$ is a causal and stationary AR($K-1$) process, the number of groups $G$ is correctly specified such that $\sigma^2_i = \kappa_g$ $\frac{1}{n_i} \mathbf{S}_i \stackrel{P}{\rightarrow} \kappa_g \zeta_g$. As $n_i \rightarrow \infty$ for all $i$, then $\hat{z}_{ig} \stackrel{P}{\rightarrow} z_{ig}$ and $\hat{\Phi}_g \stackrel{P}{\rightarrow} \Phi_g$.
\end{theorem}
The $\hat{z}_{ig}$ estimates consist of the sum of ratios of two Wishart kernels where group memberships are driven by the separation of $\zeta_j$ and $\zeta_g$. However, when we allow for a large number of groups $G$, individuals belonging to the same AR model will also be clustered by their innovation variance such that $\sigma^2_i = \kappa_g$. In practice, we use a penalization criteria such as BIC to limit the number of groups, $G$. As a result, the WMM clustering is primarily driven by differences between AR coefficients, but can also take large differences in innovation variances into consideration. In certain situations, such as forecasting, it's advantageous to group individuals by both AR coefficients and uncertainty, in order to avoid mixing large and small variances into a group level forecast model.

One can also disregard the innovation variances during clustering by using the a normalized scatter matrix $\left[ \mathbf{S}_i \right]_{rc} = n_i\hat{\gamma}_i(|r-c|) / \hat{\gamma}_i(0) = n_i \hat{\rho}_i(|r-c|)$ based on autocorrelations. The autocorrelations $\hat{\rho}_i(|r-c|)$ are also consistently estimated. As a result, we consistently estimate groups with the same AR coefficients but do not take into account heterogeneous innovation variances in the clustering algorithm.

The WMM group matrices is a means of summing weighted moments from different time series, akin to Generalized Estimating Equations and mixtures of marginal models, leading to a straightforward asymptotic distribution involving a sandwich estimator \citep{rosen2000mixtures}:

\begin{theorem} \label{thm3}
The asymptotic distributions of Yule-Walker estimates, $\hat{\Phi}_g$ derived from the Wishart mixture model are
$$
\hat{\Phi}_g \stackrel{D}{\sim} \mathrm{N}\left(\Phi_g , 
\left(\sum^I_{i=1} {z}_{ig} \mathbf{X}^T_i \mathbf{X}_i \right)^{-1}
\left(\sum^I_{i=1} {z}^2_{ig} \sigma^2_{ig} \mathbf{X}^T_i \mathbf{X}_i \right)
\left(\sum^I_{i=1} {z}_{ig} \mathbf{X}^T_i \mathbf{X}_i \right)^{-1}
\right).
$$
\end{theorem}
\noindent Often, the matrices are set to $\left[ \mathbf{X}_i^T \mathbf{X}_i \right]_{rc} = n_i\hat{\gamma}_i(|r-c|)$, replacing each element with its MLE and we used the YW estimate for variance evaluated at group $g$, $\hat{\sigma}^2_{ig} = \hat{\gamma}_i(0) \left[ 1 - \mathbf{u}^T_g \mathbf{Q}_g^{-1} \mathbf{u}_g / q_g \right]$. In addition, $\hat{z}_{ig}$, continuous probability value defined in (\ref{EM1z}) are substituted into the asymptotic distribution. The proofs are left to Appendix \ref{proof} and \ref{proof2}.

\subsection{Simulation Studies}

To evaluate the performance of competing methods we conducted two simulation studies. We denote the algorithm based on equations \eqref{EM1z}--\eqref{EM1sigma} as EM1 and the algorithm based on equations \eqref{EM3z}--\eqref{EM3lambda} as EM2. The other methods considered are the autocorrelation distance clustering (ACF), the partial autocorrelation distance clustering (PACF), the Piccolo distance clustering (PIC), the Gaussian mixture model (GMM) and the Hierarchical Spectral Merger (HSM). All autocorrelation and AR coefficient based methods were performed using two lags and all competing methods were based on two pre-specified groups. 

We simulated $I=200$ individuals with $G=2$ groups and 100 individuals in each group. We simulated a number of cases under different ARMA parameterizations, different values of $n_i$ and $\sigma^2_i$. The ARMA($p,q$) parameterization are given as 
$$
Y_i^{(t)} = \phi_g^{(1)} Y_i^{(t-1)}+ \cdots + \phi_g^{(p)} Y_i^{(t-p)} + 
\varepsilon_i^{(t)} + \theta_g^{(1)} \varepsilon_i^{(t-1)}+\cdots+\theta_g^{(q)} \varepsilon_i^{(t-q)}
$$ 
where $\varepsilon_i^{(t)} \sim \mathrm{N}(0,\sigma_i^2)$. In our simulations, after we obtain $z_{ig}$ MAP estimates we compare them to the truth and calculate accuracy of recovering the true group memberships as the number of correctly identified individuals divided by the total number of individuals. We repeated the simulation 1000 times for each case, in order to calculate mean accuracy and its standard errors (SE).


\begin{table}[H]
\caption{\textbf{Simulation Settings}. The six different scenarios considered, varying in their ARMA parameterization, sample size $n_i$ and innovation variance $\sigma^2_i$.}
\begin{center}
\scriptsize
\begin{tabular}{ |c||lclll|  }
\hline
 Case Number & Group $g$ & ARMA Model & $n_i$ & $\sigma^2_i$ &  \\
 \hline
 1 & $g=1$ & $Y_i^{(t)} = 0.6 Y_i^{(t-1)} - 0.05 Y_i^{(t-2)} + \varepsilon_i^{(t)}$ & $n_i = 100$ & 
 $\sigma^2_i=0.01$ & $1 \leq i \leq 100$\\
 & $g=2$ & $Y_i^{(t)} = 0.5 Y_i^{(t-1)} - 0.1 Y_i^{(t-2)} + \varepsilon_i^{(t)}$ & $n_i = 100$ & $\sigma^2_i=0.01$ & $101 \leq i \leq 200$\\
 \hline
 2 & $g=1$ & $Y_i^{(t)} = 0.6 Y_i^{(t-1)} - 0.05 Y_i^{(t-2)} + \varepsilon_i^{(t)}$ & $n_i = 100$ & $\sigma^2_i=100$ & $1 \leq i \leq 100$\\
 & $g=2$ & $Y_i^{(t)} = 0.5 Y_i^{(t-1)} - 0.1 Y_i^{(t-2)} + \varepsilon_i^{(t)}$ & $n_i = 100$ & $\sigma^2_i=100$ & $101 \leq i \leq 200$\\
 \hline
 3 & $g=1$ & $Y_i^{(t)} = 0.75 Y_i^{(t-1)} - 0.05 Y_i^{(t-2)} + \varepsilon_i^{(t)}$ & $n_i = 100$ & $\sigma^2_i=1$ & $1 \leq i \leq 50$ \\
 & & & $n_i = 1000$ & $\sigma^2_i=1$ & $51 \leq i \leq 100$ \\
 & $g=2$ & $Y_i^{(t)} = 0.65 Y_i^{(t-1)} - 0.1 Y_i^{(t-2)} + \varepsilon_i^{(t)}$ & $n_i = 100$ & $\sigma^2_i=1$ & $101 \leq i \leq 150$ \\
 & & & $n_i = 1000$ & $\sigma^2_i=1$ & $151 \leq i \leq 200$ \\
 \hline
 4 & $g=1$ & $Y_i^{(t)} = 0.75 Y_i^{(t-1)} - 0.05 Y_i^{(t-2)} + \varepsilon_i^{(t)}$ & $n_i = 100$ & $\sigma^2_i=1$ & $1 \leq i \leq 50$ \\
 & & & $n_i = 100$ & $\sigma^2_i = 100$ & $51 \leq i \leq 100$ \\
 & $g=2$ & $Y_i^{(t)} = 0.65 Y_i^{(t-1)} - 0.1 Y_i^{(t-2)} + \varepsilon_i^{(t)}$ & $n_i = 100$ 
& $\sigma^2_i=1$ & $101 \leq i \leq 150$ \\
 & & & $n_i = 100$ & $\sigma^2_i = 100$ & $151 \leq i \leq 200$\\
 \hline
 5 & $g=1$ & $Y_i^{(t)} = \varepsilon_i^{(t)} + 0.95 \varepsilon_i^{(t-1)}$ & $n_i=100$ & $\sigma^2_i=100$ & $1 \leq i \leq 100$\\
 & $g=2$ & $Y_i^{(t)} = \varepsilon_i^{(t)} + 0.75 \varepsilon_i^{(t-1)}$ & $n_i=100$ & $\sigma^2_i=100$ & $101 \leq i \leq 200$ \\
 \hline
 6 & $g=1$ & $Y_i^{(t)} = \varepsilon_i^{(t)} + 0.95 \varepsilon_i^{(t-1)}$ & $n_i=100$ & $\sigma^2_i=100$ & $1 \leq i \leq 50$\\
 & & & $n_i=1000$ & $\sigma^2_i=100$ & $51 \leq i \leq 100$\\
 & $g=2$ & $Y_i^{(t)} = \varepsilon_i^{(t)} + 0.75 \varepsilon_i^{(t-1)}$ & $n_i=100$ & $\sigma^2_i=100$ & $101 \leq i \leq 150$\\
 & & & $n_i=1000$ & $\sigma^2_i=100$ & $151 \leq i \leq 200$\\
 \hline
\end{tabular}
\end{center}
\label{tab:1}
\end{table}

\begin{table}[H]
\caption{\textbf{Simulated Mean Accuracy and Standard Errors of Competing Methods}. Note that for HSM, the \texttt{HSMClust R} package requires equal length time series and so could not be applied to Case 3 and 6.}
\begin{center}
\scriptsize
\begin{tabular}{ |c||ccccccc|  }
 \hline
 & \multicolumn{7}{c|}{Methods: mean accuracy (SE)} \\
 \cline{2-8}
 Case Number & ACF & PACF & PIC & GMM & HSM & EM1 & EM2 \\
 \hline
 1 & 0.711(0.058) & 0.633(0.074) & 0.576(0.055) & 0.6(0.074) & 0.587(0.101) & 0.689(0.042) & 0.69(0.043) \\
 2 & 0.716(0.056) & 0.639(0.076) & 0.577(0.056) & 0.6(0.072) & 0.585(0.1) & 0.692(0.045) & 0.692(0.044) \\
 3 & 0.781(0.126) & 0.678(0.134) & 0.589(0.088) & 0.745(0.084) & NA & 0.881(0.02) & 0.881(0.02) \\
 4 & 0.738(0.065) & 0.635(0.082) & 0.569(0.052) & 0.654(0.102) & 0.632(0.123) & 0.744(0.034) & 0.744(0.035) \\
 5 & 0.527(0.021) & 0.548(0.034) & 0.576(0.037) & 0.581(0.04) & 0.51(0.011) & 0.712(0.032) & 0.712(0.031) \\
 6 & 0.525(0.02) & 0.556(0.051) & 0.587(0.063) & 0.524(0.018) & NA & 0.838(0.023) & 0.838(0.023) \\
 \hline
\end{tabular}
\end{center}
\label{tab:2}
\end{table}

From our simulations, we found that the WMM method generally outperforms or is comparable to other methods while having lower standard errors. This is partly a result of being able to use cluster information during the model fitting due to their simultaneous fit as opposed to doing clustering sequentially after model fitting. The HSM clustering is sensitive given the small sample size but has accurate results as $n_i$ increases. ACF and WMM are comparable under the AR parameterization (Case 1--4), but are inferior to WMM under the moving average (MA) parameterization (Case 5--6). WMM are also leading methods when there are heterogeneous innovation variances (Case 4), indicating that AR coefficients drive clustering and a very large difference in innovation variance must be present in order to impact clustering. 

Under the MA parameterization, the WMM out performs all competing methods, while AR coefficient based methods suffer from model misspecification. WMM clustering does not require any ARMA modeling assumptions and works well for different types of stationary time series. Finally, WMM works well under imbalance sample sizes (Case 3,6). Time series with large $n_i$ are better able to capture the underlying process, and anchor the estimation of $\Sigma_g$ in the EM algorithm. Equation (\ref{EM1sigma}) estimates the $\Sigma_g$ as a grand mean of all scatter matrices, giving more weight to time series with large $n_i$.

\subsection{Application: COVID-19}

Numerous studies have been proposed to forecast the spread of COVID using the autoregressive integrated moving average (ARIMA) model \citep{benvenuto2020application,ceylan2020estimation,alzahrani2020forecasting}. In order to look at stationary segment of the data, we study case counts from the second winter (October 1, 2020--February 28, 2021) of the COVID pandemic. Daily new COVID cases for 67 Pennsylvania (PA) counties was obtained from the \textit{The New York Times} GitHub: \url{https://github.com/nytimes/covid-19-data} \citep{times2020covid}. Case rates were calculated by dividing daily new cases by county population and were also mean centered to create $I=67$ time series with $n_i=151$. Counties of PA differ greatly by their population offset, i.e. innovation variance, making the WMM well suited to this task. We evaluate $G=1,\cdots,10$ clusters using EM1 and BIC defined in \eqref{AIC} for selecting $G$. COVID reporting is known to be influenced by day of the week, therefore we elect to use $K-1=7$ coefficients. We initialize our WMM using HSM clustering results and we found that BIC selected for 5 groups. In addition, we compare clustering of WMM and HSM algorithms as both are designed for stationary time series.


\begin{figure}[H]
\centering
\includegraphics[width=6in]{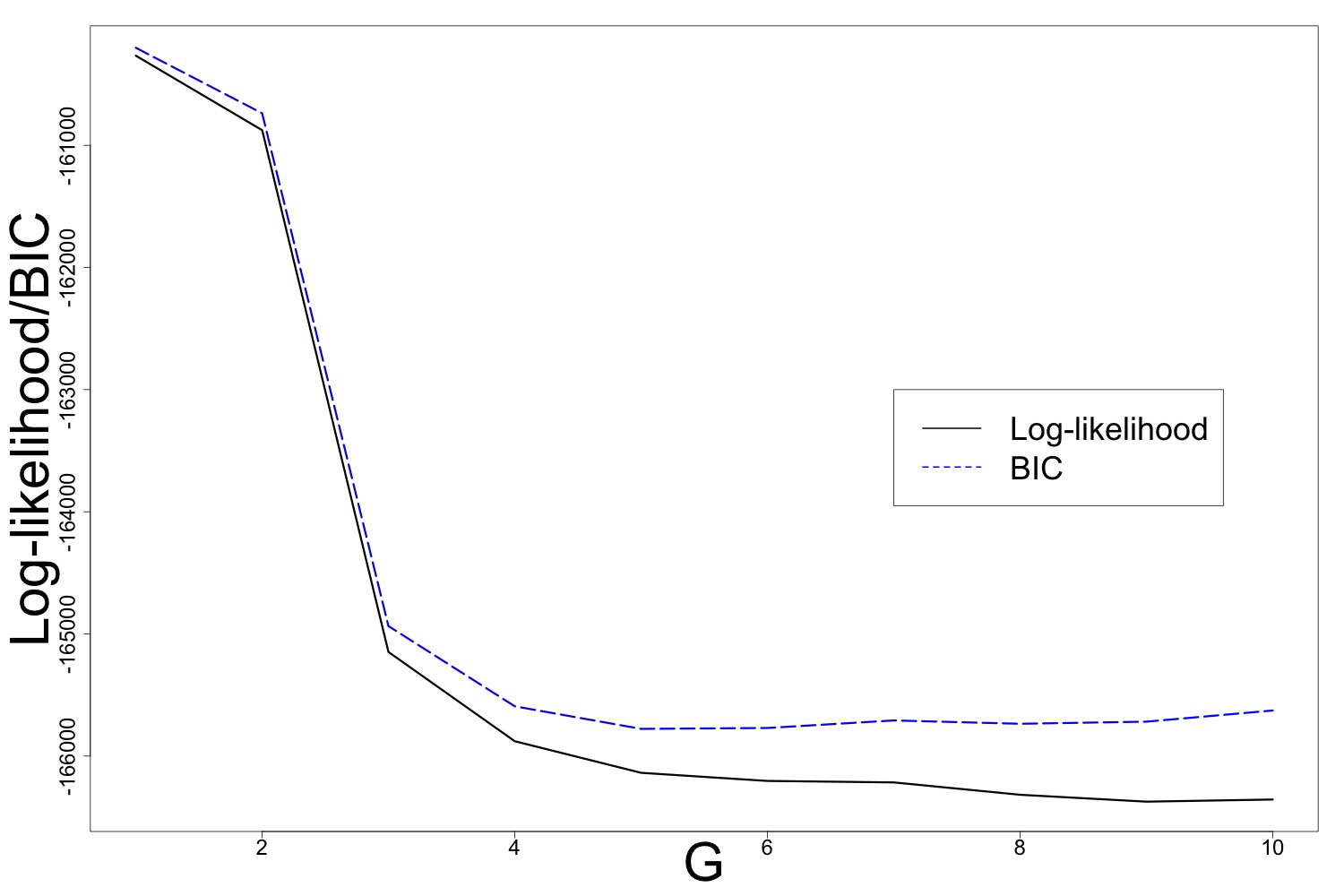}
\caption{ \textbf{BIC vs Group Number $G$}. Based on BIC, $G=5$ groups were selected.}
\label{fig:1}
\end{figure}
\normalsize

\begin{figure}[H]
\centering
\includegraphics[width=6in]{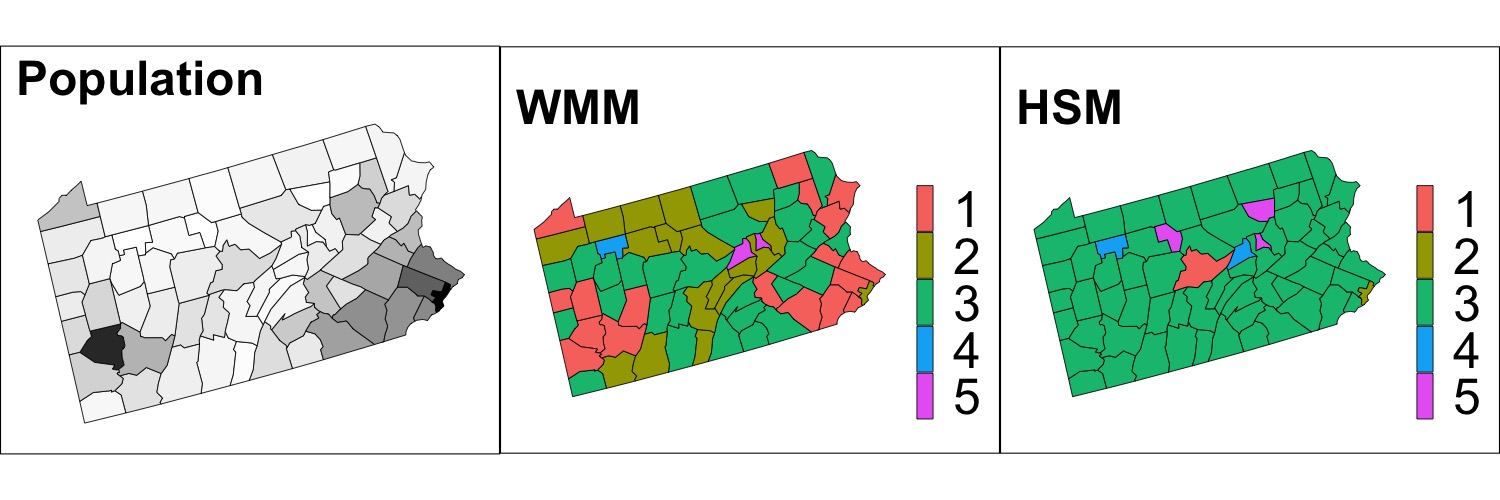}
\caption{ \textbf{Population and Estimated $z_{ig}$ of PA Counties}. WMM clustering mostly aligns with county population (gray scale, black largest population). HSM clustering tends to identify outliers first, such as sparse or densely populated counties, e.g. Philadelphia.}
\label{fig:1}
\end{figure}

\begin{table}[H]
\centering
\caption{\textbf{AR(7) Coefficient Estimates and Standard Errors}. Five sets of AR coefficients and asymptotic variances were estimated using the WMM and YW.}
\small
\begin{tabular}{|c||lllll|}
\hline 
& \multicolumn{5}{c|}{Coefficent (SE)} \\
\cline{2-6}
 & $g=1$ & $g=2$ & $g=3$ & $g=4$ & $g=5$ \\ 
  \hline \hline
  $\phi_g^{(1)}$ & 0.3347(0.0198) & 0.2(0.02) & 0.2641(0.0153) & -0.0089(0.0814) & 0.1007(0.0575) \\ 
  $\phi_g^{(2)}$ & 0.1493(0.0206) & 0.086(0.0202) & 0.1449(0.0157) & -0.0332(0.0812) & -0.0016(0.0577) \\ 
  $\phi_g^{(3)}$ & -0.0067(0.0208) & 0.0683(0.0203) & 0.0773(0.0158) & 0.233(0.0811) & 0.1637(0.0574) \\ 
  $\phi_g^{(4)}$ & 0.0864(0.0208) & 0.0498(0.0203) & 0.0615(0.0158) & 0.0013(0.0833) & 0.0054(0.0579) \\ 
  $\phi_g^{(5)}$ & -0.0238(0.0208) & 0.0888(0.0203) & 0.0031(0.0158) & 0.062(0.0811) & 0.1247(0.0574) \\ 
  $\phi_g^{(6)}$ & 0.1642(0.0206) & 0.144(0.0202) & 0.1451(0.0157) & -0.0661(0.0812) & 0.0321(0.0577) \\ 
  $\phi_g^{(7)}$ & 0.2192(0.0198) & 0.1937(0.02) & 0.2035(0.0153) & -0.0107(0.0814) & 0.0993(0.0575) \\ 
   \hline
\end{tabular}
\label{tab:3}
\end{table}

When using HSM, Philadelphia was first separated into its own group, follow by the rural counties. Under HSM, the majority of the PA county belong to the same group and results are not meaningful with many clusters being occupied by singletons. This is because distance-based algorithms often separate outliers are the start, while mixture model based methods are more robust because they seek to initially estimate group parameters, which leads to more balanced groups. 

Under WMM, Group 1 ($g=1$) is comprised of suburbs and populous counties, with the exception of Philadelphia. The reason for this is because Philadelphia uniquely does not report cases on the weekends, leading to weaker apparent autocorrelations and clustering Philadelphia instead with rural counties ($g=2$). Sparsely populated counties tend to have low autocorrelations, while autocorrelation tends to be positively correlated with county population. Populous counties have regimented testing protocols leading to higher autocorrelation. Our method simultaneously clustered and estimated AR models for each group. Combined with the asymptotic distribution for evaluating our coefficients, our simplified procedure is a fast and intuitive method for studying heterogeneous time series data. Our AR models serve as a parsimonious forecast models that borrow information across different counties separated into meaningful clusters.

\section{Conclusion}

We proposed a computationally efficient method to cluster stationary time series and estimate their group AR model. Our method incorporates different innovation variances and sample sizes in the estimation, making it suitable for heterogeneous datasets. Under mild conditions our AR models and group labels estimates are consistent and have asymptotic distributions which accounts for heterogeneous variances. From simulations, we found that our WMM approach outperforms most competing methods. Furthermore, our WMM approach improves as sample size increases even in datasets with imbalanced time series lengths. 

Our WMM and ARMM shows promise as a clustering and forecasting model for COVID cases in PA counties. We found that group assignments mostly align with county population, and PA COVID time series primarily consist of three main groups with different levels of autocorrelations and distinct AR models. As a future analysis, we may incorporate the detrending procedure of \cite{qiu2013efficient} in order to study non-stationary time series. In conclusion, our WMM method is well equipped to handle noisy and large datasets by efficiently combining clustering with model fitting in a mixture model framework.

%
%

\bigskip
\begin{center}
{\large\bf APPENDIX}
\end{center}

\section{Proof of Theorem \ref{thm2}} \label{proof}

Since we evaluate the estimator as all $n_i \rightarrow \infty$, for ease of notation, assume $n_i=n\ \forall i$ and $n \rightarrow \infty$. In addition, assume that $\mathbf{y}_i$ has mean zero. By the asymptotic behavior of the Yule-Walker estimates for causal AR processes (Theorem \ref{thm1}), scatter matrices as a sequence of $n$, converges in probability to their correct group autocovariance matrix, $\mathbf{S}_{i}/n \stackrel{P}{\rightarrow} \kappa_g \zeta_g$, $\hat{\sigma}_i^2 \stackrel{P}{\rightarrow} \kappa_g$, $g=1,2,3,\dots,G$, and $G \le I$. Evaluating $\hat{\Sigma}_{g}$ 
$$
\hat{\Sigma}_{g} = \frac{\sum_{i=1}^{I} \hat{z}_{i g} \mathbf{S}_{i}}{ n\sum_{i=1}^{I} \hat{z}_{i g}}
$$
suppose labels $z_{i g}$ are correct for group $g$, $\hat{\Sigma}_{g}$ then $\Sigma_{g} = \kappa_g \zeta_g = \frac{\sum^I_{i=1} z_{ig} \sigma^2_i}{\sum^I_{i=1} z_{ig}} \zeta_g$. Update for the group indicator, $\hat{z}_{i g} = \operatorname{Pr} \left( Z_{i g}=1 | \mathbf{S}_{i}, \hat{\Theta} \right)$, are given as 
\begin{equation} \label{EMz}
\hat{z}_{i g}
=
\frac{\pi_{g} f_{W}\left(\mathbf{S}_{i} | \Sigma_{g}, n\right)}
{\sum_{g=1}^{G} \pi_{g} f_{W}\left(\mathbf{S}_{i} | \Sigma_{g}, n\right)} .
\end{equation}
The pdf of the Wishart distribution is given as:
\begin{align*}
f_{W}\left( \mathbf{S}_{j} | \Sigma_{g}, n\right)
=& 
\underbrace{
\frac{\left| \mathbf{S}_{j} \right|^{-(K+1) / 2}}{\pi^{K(K-1) / 4}}
}_{M_1} 
\underbrace{
\left|\frac{ \mathbf{S}_{j} \Sigma_{g}^{-1}}{2}\right|^{n / 2} 
}_{A}
\underbrace{
\exp \left( \frac{-1}{2} \operatorname{trace} \left( \Sigma_{g}^{-1} \mathbf{S}_{j} \right) \right)
}_{B}
\underbrace{
\frac{1}{\prod_{k=1}^{K} \Gamma\left(\frac{n-k+1}{2}\right)}
}_{C}
\end{align*}
where $M_1$ is constant with respecct to $\Sigma_g$, but tends to 0 as $n$ gets large. However, $M_1$ is factored out and cancels, in the numerator and denominator, when evaluating $z_{ig}$, (\ref{EMz}).

When the individual is in the correct group, then $\Sigma_{g}^{-1} \mathbf{S}_{i} = n \mathbf{I} $, $A=\left( \frac{n}{2}\right)^{K n/2}$, $B=\exp \left(\frac{- n K}{2}\right)$ and $AB = \left(\left(\frac{ n }{2 e }\right)^{n / 2}\right)^{K}$. Evaluating the lower bound for $C$ at $k=1$,
$$
C >\frac{1}{\prod_{k=1}^{K} \Gamma\left(\frac{n}{2}\right)} =\left(\frac{1}{\Gamma\left(\frac{n}{2}\right)}\right)^{K}
$$
and the lower bound of $ABC$ is given as
$$
ABC>\left(\frac{\left(\frac{n}{2 e }\right)^{n / 2}}{\Gamma\left(\frac{n}{2}\right)}\right)^{K}=D^{K}.
$$
Using the Laplace method or Stirling's formula for gamma function,
$$
\Gamma(t)=\sqrt{\frac{2 \pi}{t}}\left(\frac{t}{e}\right)^{t}\left(1+O\left(\frac{1}{t}\right)\right)
$$
to evaluate $D$, we get $D = \sqrt{ \frac{n}{4 \pi } } / \left( 1+O\left(\frac{1}{n}\right) \right) = O\left( \sqrt{n} \right)$, when $\mathbf{S}_{i}/n \stackrel{P}{\rightarrow} \kappa_g \zeta_g$ and we have $ABC = O\left( n^{K/2} \right)$.

When $\mathbf{S}_i/n \stackrel{P}{\rightarrow} \sigma^2_i \zeta_g$, the individual is in the incorrect group, then $\Sigma_{j}^{-1} \mathbf{S}_{i} = n \kappa^{-1}_j \sigma^2_i \zeta^{-1}_j \zeta_g = n \kappa^{-1}_j \sigma^2_i \mathbf{W}_{jg}$, $m_k(\mathbf{W}_{jg})$ are the eigenvalues of $\mathbf{W}_{jg}=\zeta^{-1}_j \zeta_g$. The rest is given as $A=\left(\frac{n}{2}\right)^{K n / 2}\left(\prod_{k=1}^{K} \kappa^{-1}_j \sigma^2_i m_k(\mathbf{W}_{jg}) \right)^{n / 2}$, and $B= \exp \left( \sum_{k=1}^{K} \kappa^{-1}_j \sigma^2_i m_k(\mathbf{W}_{jg}) \right)^{-n / 2}$. Evaluating the upper bound for $C$ at $k=K$, $C< \left(\frac{1}{\Gamma\left(\frac{n-K+1}{2}\right)}\right)^{K}$. Thus, the upper bound for $ABC$ is given as
\begin{align*}
ABC &<\left(\frac{n}{2}\right)^{K n / 2}\left(\prod_{k=1}^{K} \kappa^{-1}_j \sigma^2_i m_k(\mathbf{W}_{jg}) \right)^{n / 2}\left(\frac{1}{\exp \left( \sum_{k=1}^{K} \kappa^{-1}_j \sigma^2_i m_k(\mathbf{W}_{jg}) \right) }\right)^{n / 2}\left(\frac{1}{\Gamma\left(\frac{n-K+1}{2}\right)}\right)^{K} \\
&=
\underbrace{
\left(\frac{\left(\frac{n}{2}\right)^{n / 2}}{\Gamma\left(\frac{n-K+1}{2}\right)}\right)^{K}
}_{E^{K}}
\left( \prod_{k=1}^{K} \frac{ \kappa^{-1}_j \sigma^2_i m_k(\mathbf{W}_{jg}) }{\exp \left( \kappa^{-1}_j \sigma^2_i m_k(\mathbf{W}_{jg}) \right)}\right)^{n / 2}
\end{align*}
Using Stirling's formula to evaluate $E$, we get
\begin{align*}
E &=\frac{\left(\frac{n}{2}\right)^{n / 2}}{\Gamma\left(\frac{n-K+1}{2}\right)} \\
&=\frac{\left(\frac{n}{2}\right)^{n / 2}}{\sqrt{\frac{4 \pi}{n-K+1}}\left(\frac{n-K+1}{2 e}\right)^{\frac{n-K+1}{2}}\left(1+O\left(\frac{1}{n}\right)\right)} \\
&=\frac{\left(\frac{n}{2}\right)^{n / 2} \sqrt{n-K+1}}{\sqrt{4 \pi}\left(\frac{n-K+1}{2 e}\right)^{\frac{n-K+1}{2}}\left(1+O\left(\frac{1}{n}\right)\right)} \\
&=\frac{\left(\frac{n}{2}\right)^{n / 2} \sqrt{n-K+1}\left(\frac{2 e}{n-K+1}\right)^{\frac{n-K+1}{2}}}{\sqrt{4 \pi}\left(1+O\left(\frac{1}{n}\right)\right)} \\
&=\frac{\left(\frac{n}{2}\right)^{n / 2} \sqrt{n-K+1}\left(\frac{2 e}{n-K+1}\right)^{\frac{n}{2}}\left(\frac{2 e}{n-K+1}\right)^{\frac{-K+1}{2}}}{\sqrt{4 \pi}\left(1+O\left(\frac{1}{n}\right)\right)} \\
&=\frac{\sqrt{n-K+1} e^{n / 2}\left(\frac{n}{n-K+1}\right)^{\frac{n}{2}}\left(\frac{2 e}{n-K+1}\right)^{\frac{-K+1}{2}}}{\sqrt{4 \pi}\left(1+O\left(\frac{1}{n}\right)\right)} \\
&=\frac{\sqrt{n-K+1} e^{n / 2}\left(\frac{n}{n-K+1}\right)^{\frac{n}{2}}\left(\frac{2}{n-K+1}\right)^{\frac{-K+1}{2}} e^{\frac{-K+1}{2}}}{\sqrt{4 \pi}\left(1+O\left(\frac{1}{n}\right)\right)} \\
&=\frac{\sqrt{n-K+1} e^{n / 2}\left(\frac{n}{n-K+1}\right)^{\frac{n}{2}}\left(\frac{n-K+1}{2}\right)^{\frac{K-1}{2}} e^{\frac{-K+1}{2}}}{\sqrt{4 \pi}\left(1+O\left(\frac{1}{n}\right)\right)}
\end{align*}
Because we are working in asymptotic behavior, $n \rightarrow \infty$, we use the following limit representation of $e$, 
$$
e^{-K+1} = \left(1+(-K+1) \frac{1}{n}\right)^{n}=\left(\frac{n-K+1}{n}\right)^{n} \implies 
e^{\frac{-K+1}{2}} = \left(\frac{n-K+1}{n}\right)^{ \frac{n}{2} }.
$$
Continuing the derivation for $E$, we get
\begin{align*}
E &=\frac{\sqrt{n-K+1} e^{n / 2}\left(\frac{n}{n-K+1}\right)^{\frac{n}{2}}\left(\frac{n-K+1}{2}\right)^{\frac{K-1}{2}}\left(\frac{n-K+1}{n}\right)^{ \frac{n}{2} }}
{\sqrt{4 \pi}\left(1+O\left(\frac{1}{n}\right)\right)} \\
&=\frac{\sqrt{n-K+1} e^{n / 2}\left(\frac{n-K+1}{2}\right)^{\frac{K-1}{2}}}{\sqrt{4 \pi}\left(1+O\left(\frac{1}{n}\right)\right)} \\
&=\frac{(n-K+1)^{1 / 2} e^{n / 2}(n-K+1)^{\frac{K-1}{2}}}{\sqrt{4 \pi}\left(1+O\left(\frac{1}{n}\right)\right) 2^{\frac{K-1}{2}}} \\
&=\frac{e^{n / 2}(n-K+1)^{\frac{K}{2}}}{\sqrt{4 \pi}\left(1+O\left(\frac{1}{n}\right)\right) 2^{\frac{K-1}{2}}} \\
&=O\left( e^{n / 2} n^{K / 2} \right)
\end{align*}
The upper bound of $ABC$ is now given as
$$
ABC < \left( e^{n / 2} n^{K / 2} \right)^K \left( \prod_{k=1}^{K} \frac{ \kappa^{-1}_j \sigma^2_i m_k(\mathbf{W}_{jg}) }{\exp \left( \kappa^{-1}_j \sigma^2_i m_k(\mathbf{W}_{jg}) \right)}\right)^{n / 2}
= n^{M_2} \left( \prod_{k=1}^{K} \frac{ \kappa^{-1}_j \sigma^2_i m_k(\mathbf{W}_{jg}) }{\exp \left( \kappa^{-1}_j \sigma^2_i m_k(\mathbf{W}_{jg}) - 1 \right)} \right)^{n / 2}
$$
where $M_2$ is a constant with respect to $n$. Finally, $\frac{ x }{\exp \left( x - 1 \right)} \le 1$ and the equality only holds when $x=1$, but in our case $\kappa^{-1}_j \sigma^2_i m_k(\mathbf{W}_{jg}) \ne 1$ for every $k$ because $\kappa^{-1}_j \sigma^2_i \mathbf{W}_{jg} \ne \mathbf{I}$. Therefore, $0 < r_{ig} = \prod_{k=1}^{K} \frac{  \kappa^{-1}_j \sigma^2_i m_k(\mathbf{W}_{jg}) }{\exp \left( \kappa^{-1}_j \sigma^2_i m_k(\mathbf{W}_{jg}) - 1 \right)} < 1$. The upper bound for $ABC$ is given as $O \left( n^{M_2} \left( \prod_{k=1}^{K} \frac{ \kappa^{-1}_j \sigma^2_i m_k(\mathbf{W}_{jg}) }{\exp \left( \kappa^{-1}_j \sigma^2_i m_k(\mathbf{W}_{jg}) - 1 \right)} \right)^{n/2} \right)$, when $\mathbf{S}_i/n \stackrel{P}{\rightarrow} \sigma^2_i \zeta_g$. If $\mathbf{S}_i/n \stackrel{P}{\rightarrow} \sigma^2_i \zeta_j$ we have $0 < \prod_{k=1}^{K} \frac{  \kappa^{-1}_j \sigma^2_i }{\exp \left( \kappa^{-1}_j \sigma^2_i  - 1 \right)} < 1$, when $\sigma^2_i \ne \kappa_j$, in other words clusters must also have the same AR coefficients and innovation variances.

When the individual is in the correct group $\mathbf{S}_{i}/n \stackrel{P}{\rightarrow} \kappa_g \zeta_g$, and $\sigma_i^2 = \kappa_g$, then $\hat{z}_{i g} = \operatorname{Pr} \left( Z_{i g}=1 | \mathbf{S}_{i}, \hat{\Theta} \right)$ becomes
$$
\hat{z}_{i g}
\geq
\frac{ \pi_{g} O(n^{K/2}) }
{\pi_{g} O(n^{K/2}) + \sum_{j\neq g} \pi_{j} O \left( n^{M_2}r_{jg}^{n/2} \right) }
$$
where $0<r_{jg}<1$. Then consistency of $\hat{z}_{i g}$ follows, $\lim _{n \rightarrow \infty} \operatorname{Pr}\left(\left|\hat{z}_{i g}-{z}_{i g}\right|>\varepsilon\right)=0$. 

Under strict concavity, iterative convergence of the EM algorithm reaches the maximum, our label given in \eqref{EMz} converge in probability to the indicator function for the correct group. We now have $\hat{z}_{i g} \stackrel{P}{\rightarrow} z_{ig}$, $\mathbf{S}_{i}/n \stackrel{P}{\rightarrow} \kappa_g \zeta_g$ and
$$
\hat{\Sigma}_{g} = \frac{\sum_{i=1}^{I} \hat{z}_{i g} \mathbf{S}_{i}}{ n\sum_{i=1}^{I} \hat{z}_{i g}}
\stackrel{P}{\rightarrow} \kappa_g \zeta_g .
$$
$\hat{\Sigma}_g = 
\left[\begin{array}{c|c}
q_g & \mathbf{u}_g^T  \\
\hline
\mathbf{u}_g & \mathbf{Q}_g
\end{array}\right]$ is a consistent AR coefficient estimator through Yule-Walker equations because it only relies on proportionality to $\zeta_g$. We have $\hat{\Phi}_g = \mathbf{Q}_g^{-1} \mathbf{u}_g$, and $\hat{\Phi}_g \stackrel{P}{\rightarrow} \Phi_g$.

\section{Proof of Theorem \ref{thm3}} \label{proof2}
From Theorem \ref{thm1} we have 
$$
(\mathbf{X}^T_i \mathbf{X}_i )^{-1} \mathbf{X}^T_i \mathbf{y}_i = \hat{\Phi}_g \stackrel{D}{\sim} \mathrm{N}\left(\Phi_g, \sum \sigma_i^{2} (\mathbf{X}^T_i \mathbf{X}_i )^{-1}\right)
$$
therefore 
\begin{align*}
\mathbf{X}^T_i \mathbf{y}_i  &\stackrel{D}{\sim} \mathrm{N} \left( \mathbf{X}^T_i \mathbf{X}_i \Phi_g, \sigma_i^{2} \mathbf{X}^T_i \mathbf{X}_i \right) \\
\sum^I_{i=1} z_{ig} \mathbf{X}^T_i \mathbf{y}_i 
&\stackrel{D}{\sim} \mathrm{N} \left( \left(\sum^I_{i=1} {z}_{ig} \mathbf{X}^T_i \mathbf{X}_i \right) \Phi_g, \left(\sum^I_{i=1} {z}^2_{ig} \sigma^2_i \mathbf{X}^T_i \mathbf{X}_i \right) \right) \\
\left(\sum^I_{i=1} {z}_{ig} \mathbf{X}^T_i \mathbf{X}_i \right)^{-1}
\sum^I_{i=1} z_{ig} \mathbf{X}^T_i \mathbf{y}_i
&\stackrel{D}{\sim} \mathrm{N} \left( \Phi_g, \left(\sum^I_{i=1} {z}_{ig} \mathbf{X}^T_i \mathbf{X}_i \right)^{-1} \left(\sum^I_{i=1} {z}^2_{ig} \sigma^2_i \mathbf{X}^T_i \mathbf{X}_i \right) \left(\sum^I_{i=1} {z}_{ig} \mathbf{X}^T_i \mathbf{X}_i \right)^{-1} \right) 
\end{align*}
and $\left(\sum^I_{i=1} {z}_{ig} \mathbf{X}^T_i \mathbf{X}_i \right)^{-1}
\sum^I_{i=1} z_{ig} \mathbf{X}^T_i \mathbf{y}_i = \hat{\Phi}_g$
$$
\hat{\Phi}_g \stackrel{D}{\sim} \mathrm{N} \left( \Phi_g, \left(\sum^I_{i=1} {z}_{ig} \mathbf{X}^T_i \mathbf{X}_i \right)^{-1} \left(\sum^I_{i=1} {z}^2_{ig} \sigma^2_i \mathbf{X}^T_i \mathbf{X}_i \right) \left(\sum^I_{i=1} {z}_{ig} \mathbf{X}^T_i \mathbf{X}_i \right)^{-1} \right).
$$
Note that $\sum^I_{i=1} {z}_{ig} \mathbf{X}^T_i \mathbf{X}_i / M_3 = \mathbf{Q}_g$ and $\sum^I_{i=1} z_{ig} \mathbf{X}^T_i \mathbf{y}_i / M_3 = \mathbf{u}_g$, where $M_3$ is a common denominator.

\bibliographystyle{JASA}

\bibliography{WMM_manuscript_preprint}
\end{document}